\documentclass[onecolumn]{emulateapj}

\usepackage{graphicx}

\renewcommand{\vec}[1]{\mathbf{#1}}
\newcommand{\levy}{L\'{e}vy }

\begin{document}

\title{A Unified Theory for the Effects of Stellar Perturbations and Galactic Tides
on Oort Cloud Comets}
\author{Benjamin F. Collins\altaffilmark{1} and Re'em Sari\altaffilmark{1,2}}
\altaffiltext{1}{California Institute of Technology, MC 130-33,
        Pasadena, CA 91125}
\altaffiltext{2}{Racah Institute of Physics, Hebrew University, Jerusalem 91904,
Israel}
\email{bfc@tapir.caltech.edu}

\begin{abstract}
We examine the effects of passing field stars on the angular momentum
of a nearly radial orbit of an Oort cloud comet bound to the Sun.  We
derive the probability density function (PDF) of the change in angular
momentum from one stellar encounter, assuming a uniform and isotropic
field of perturbers.  We show that the total angular momentum follows
a \levy flight, and determine its distribution function.  If there is
an asymmetry in the directional distribution of perturber velocities,
the marginal probability distribution of each component of the angular
momentum vector can be different.  The constant torque attributed to
Galactic tides arises from a non-cancellation of perturbations with an
impact parameter of order the semimajor axis of the comet.  
%The most
%appropriate model is determined by the impact parameters that are the
%most important.  
When the close encounters are rare, the angular momentum is best
modeled by the stochastic growth of stellar encounters.  If
trajectories passing between the comet and sun occur frequently, the
angular momentum exhibits the coherent growth attributed to the
Galactic tides.
\end{abstract}


\section{Introduction}

In the same work as he proposed the existence of a large reservoir of
comets in the outskirts of the solar system, \citet{O50} suggested a
two-stage process for the creation of such a cloud.  First,
perturbations from the planets increase the semimajor axes of nearby
smaller objects.  These interactions leave the periapses of the small
bodies in the planetary region, but will eventually deliver enough
energy to eject them from the solar system.  The second stage of Oort
cloud formation requires that perturbations external to the solar
system deliver angular momentum to the comets.  This raises their
periapses out of the realm of planetary influence and saves them from
eventual ejection.  Further perturbations are necessary to lower their
periapses again so that they can return to the planetary region and be
observed from Earth.

Oort's original suggestion for both the circularization and delivery
mechanisms is the influence of other stars in the Galaxy as they
encounter the solar system.  Each star that passes the solar system
delivers a small kick to each comet that depends on the mass of the
star, its velocity, and its distance of closest approach.
\citet{HT86} explored the effects of a large scale planar symmetry in
the swarm of stellar perturbers to find a smooth torque similar in
magnitude to or even dominant over the stochastic stellar
perturbations.  This effect is known as the ``Galactic tidal torque,''
since it can be attributed to the gradient of the average potential of the
Galactic disk.  Several groups have used numerical simulations to
investigate the formation of the Oort cloud from a combination of
stellar perturbations and Galactic tides \citep{DQT87,DWLD04}; in all cases the two
effects have been implemented separately.

Recent studies have provided analytic solutions to several other
stochastic scattering problems that arise in orbital dynamics and
planet formation \citep{CS06,CSS07}.  \citet{CS08} investigated the
evolution of an initially circular orbit interacting impulsively with
unbound perturbers.  They showed that the probability per unit time of
perturbing a circular orbit to an eccentricity of order $e$ is
proportional to $e^{-1}$.  This power law is enough to determine that
the eccentricity of the binary diffuses as a \levy process, and the
scale of the distribution grows linearly with time \citep{SZF95}. Such
evolution is fast compared to the common Brownian motion-type
diffusion where the distribution evolves as only the square-root of
time.

In this work, we apply the framework developed for perturbations
around nearly circular orbits in \citet{CS08}, hereafter CS08, to the
case of perturbations around nearly radial orbits.  Section
\ref{secSingleEnc} presents the effects of a single stellar passage on
a zero angular momentum comet.  In Section \ref{secLevy} we derive and
solve the Boltzmann equation that describes the accumulation of the
changes in angular momentum from an isotropic distribution of
perturbers.  Section \ref{secGalacticTides} describes the
perturbations that arise from an anisotropic velocity distribution,
and explains the connection between the effects of stellar encounters
and the tidal force from the Galactic potential.  Section
\ref{secConclusions} summarizes our conclusions.

\section{A Single Stellar Passage}
\label{secSingleEnc}

In this section we discuss the change in angular momentum,
eccentricity, and periapse of a comet on a nearly radial orbit.  We
call the central body of the system the ``sun.''  We denote the
position of the comet as $\vec r_b(t)$, and its velocity $\vec
v_b(t)$.  We write the magnitude of $\vec r_b(t)$ as $r_b(t)=|\vec
r_b(t)|$, and the unit vector as $\hat r_b(t)=\vec
r_b(t)/r_b(t)$.  Since a radial orbit is by definition a straight
line, $\hat r_b(t)$ is constant in time.  Furthermore, the direction
of the velocity, $\hat v_b(t)$, is either aligned or anti-aligned with
$\hat r_b$.  The orbital energy per unit mass of the comet, ${\cal E}$, sets
the semimajor axis, $a$, and the orbital period, $T_{\rm orb}$.  The
angular momentum vector, $\vec J$ is zero, and the eccentricity vector is 
then given by $\vec e = \vec v_b \times \vec J/(G M_{\odot}) - \hat r_b = -\hat
r_b$. Finally, determining the position of the comet as a function of
time requires specifying the time that the comet passes through
periapse, $\tau$.

We call each perturber a ``star,'' and write the velocity of the star
$\vec v_p$.  The mass of the star, $m_p$, will typically be about
the same magnitude as $M_{\odot}$, the mass of the sun; both are
very large compared to the mass of the comet, $m_c$.  We focus this
analysis on the regime where the path of the star is unaffected by the
gravity of the Sun, or $ G M_{\odot}/(b v^2_p) \ll 1$.  Then the
position of the perturber as a function of time is given by 
$\vec r_p(t) = \vec b + \vec v_p (t-t_0)$, where $\vec b$ 
describes the closest position of the star relative to the sun and 
$t_0$ is the time at which the star reaches this position.

We consider, at first, encounters between the star and the sun that
occur with $b \gg 2 a$, such that the perturbation to the sun-comet
system can be treated in the tidal limit.  We will show that these 
interactions are important for setting the angular momentum
distribution when it is near zero.  In section 4 we derive the
evolution of the angular momentum as it evolves under all types of
encounters including $b \leq a$.

In this case 
the tidal acceleration as a function of time is given by:

\begin{equation}
\label{eqAtidal}
\vec a_T(t') = G m_p r_b(t') 
\left[ \frac{\hat v_p (\hat r_b \cdot \hat v_p) - \hat r_b}
{(b^2+(v_p t')^2)^{3/2}} - 3 \frac{(\vec b + \vec v_p t)(\vec b \cdot \hat r_b)}
{(b^2 + (v_p t')^2)^{5/2}} \right],
\end{equation}

\noindent
where we have translated the time coordinate by $t_0$ 
to simplify the expression.

The acceleration caused by each passing star affects the shape of the
comet's orbit.  Since the angular momentum is initially zero, the
small impulses have a large relative effect on $\vec J$.  In contrast,
single perturbations to $\vec e$ and $a$ are always small compared to
their initial magnitudes.  The periapse of the comet,
which is important for determining the influence of the planets on the 
comet, is related to the angular momentum, $J=\sqrt{2 G M_{\odot}
  q}$.  For these reasons, we focus on understanding the effects of
the stellar perturbations on the angular momentum vector.

To find the total change in angular momentum for one stellar passage,
we integrate the acceleration over the motion of the star and of the
comet: $\Delta \vec J = \int \vec r_b(t') \times \vec a_T(t') d t'$.
There are two limiting cases where we can evaluate this integral to
find a closed form solution. 
The first is the impulsive regime, where
$b/v_p \ll T_{\rm orb}$.  
The comet spends most of its time with $r_b \sim a$; however,
for rare interactions that occur when $r_b(t) \ll a$,
impulsiveness requires $b/v_p \ll r_b(t)/v_b(t)$.
We treat the comet as
stationary over the duration of an impulsive perturbation:
$r_b(t')=r_b(t_0)$, and find the change in angular momentum to
be:

\begin{equation}
\label{eqDJImpulsive}
\Delta \vec J(t_0) =
\frac{2 G m_p r_b(t_0)^2}{v_p b^2} 
\left[ (\hat r_b \times \hat v_p)(\hat r_b \cdot \hat v_p)
- 2 (\hat r_b \times \hat b)(\hat r_b \cdot \hat b)
 \right].
\end{equation}

The other simplifying case is a very non-impulsive 
encounter ($b/v_p \gg T_{\rm orb}$).  When each orbit is very
short relative to the timescale of the perturbation, 
the acceleration at each point along the perturber's 
path is experienced by the entire span of the comet's orbit.
The disparate timescales in this regime allows the integral 
over the motion of the comet to be separated from the integral 
over the path of the star.  The result is
a $\Delta \vec J$ that is independent of $t_0$ and $\tau$:

\begin{equation}
\label{eqDJNonImpulsive}
\Delta \vec J = \frac{5}{2}
\frac{G m_p}{v_p} \left(\frac{a}{b}\right)^2
\left[ (\hat r_b \times \hat v_p)(\hat r_b \cdot \hat v_p)
- 2 (\hat r_b \times \hat b)(\hat r_b \cdot \hat b)
\right].
\end{equation}

\noindent
This result also follows from replacing  $r_b(t_0)^2$
in Equation \ref{eqDJImpulsive} with its time averaged value,
$\langle r_b^2 \rangle = (5/2)a^2$.

\section{L\'{e}vy Flight Behavior}
\label{secLevy}

Successive perturbations cause the angular momentum delivered to the
comet to accumulate.  Individual perturbations add to the existing
angular momentum vectorially: $\vec J_{\rm new} = \vec r_b \times
(\vec v_b + \vec \Delta v) = \vec J + \Delta \vec J$.  Holding $\hat
r_b$ constant restricts the angular momentum vector to a plane.  We
accordingly treat $\vec J$ as a two-dimensional vector throughout this
work.  Since the perturbations by passing stars occur randomly, we
employ the same statistical approach as CS08.  We study the evolution
of $\vec J$ by deriving a distribution function, $f(\vec J,t)$, that
specifies the probability that the comet will have an angular momentum
within the region $d^2 \vec J$ around $\vec J$ at time $t$.  If we
assume that the perturbations occur isotropically, there is no
preferred direction for the accumulated angular momentum of the comet.
We then expect that $f(\vec J,t)=f(J,t)$.  The probability of finding the
comet's angular momentum with a magnitude between $J$ and $d J$ in any
direction is $2 \pi f(J,t) J dJ$.  We relax the assumptions of 
isotropy in Section \ref{secGalacticTides}.

We express the probability density function (PDF) for single perturbations as 
a frequency per unit angular
momentum, ${\cal R}(J')$.  This function describes the probability
per unit time that the comet receives a perturbation with a magnitude between
$J'$ and $J'+d J'$.  Given the properties of the ensemble of perturbing
stars, we compute the frequency with the following expression:

\begin{equation}
\label{eqRofe}
{\cal R}(J')= \int \delta(| \Delta \vec J(\vec v_p,\vec b,t_0,m_p)| - J')
{\cal F}(\vec v_p,m_p) v_p \delta(\vec b \cdot \hat v_p) 
%\delta(t_0-t) d t_0 d(t/T_{\rm orb})
 d^3 \vec b d^3 \vec v_p dm_p d (t_0/T_{\rm orb}).
\end{equation}

\noindent
where the function ${\cal F}(\vec v_p,m_p)$ is the combined phase
space density of perturbers in $\vec v_p$ and $m_p$, normalized such
that the total mass density of perturbers in real space is $\rho =
\int m_p {\cal F}(\vec v_p,m_p) d^3 \vec v_p d m_p$.  This equation is
analogous to Equation 9 of CS08, and is a precise
formulation of the idea that the frequency at which the comet is
perturbed by an amount of order $J'$ is calculated by $J' {\cal R}(J')
\sim n v b^2$, where $n$ is the number density of perturbers, $v$ is
the velocity at which they encounter the sun-comet system, and $b^2$
is the cross-sectional area for such an encounter.  
In words, Equation \ref{eqRofe} integrates over the entire 
parameter space of the encounter geometry ($\vec v_p, \vec b, 
t_0$, and $m_p$), weights the integral by the probability
density of each parameter, and uses the delta function 
of $|\Delta J(\vec v_p, \vec b, t_0,m_p)|$ to select those 
geometries that produce a perturbation of size $J'$.

%In this context,
%we include an integral over $d(t_0/T_{\rm orb})$ to take into account
%the probability of finding the comet at a particular point in its
%orbit. 

The frequency of perturbations is linked to the distribution function
through a Boltzmann equation:

\begin{equation}
\label{eqBoltzmann}
\frac{\partial f(J,t)}{\partial t}
= \int p(\vec J') [ f(|\vec J' + \vec J|)-f(J) ] d^2 \vec J'.
\end{equation}

\noindent
As in CS08, the function $p(\vec J')$ describes the frequency per unit
angular momentum space ($d^2 \vec J'$) at which a comet with angular
momentum $\vec J$ is perturbed to $\vec J+\vec J'$; this is the PDF of
$\vec J'$.  We expect this frequency to depend only on the magnitude
of the perturbation and not the direction, $p(\vec J')=p(J')$, for
isotropic perturbers.  It is related to ${\cal R}(J')$ by integrating
$p(J')$ over the angular component of $\vec J'$, ${\cal R}(J')= 2 \pi
J' p(J')$.

We assume that the stellar perturbers have only one mass, $m_p$, and
one velocity, $v_p$, that can point in any direction.  The calculation
of $p(J')$ then proceeds similarly to the calculation presented in
CS08.  Since the angular momentum excited by a perturber is
proportional to $m_p$, $v_p$, and $b$ in all the same ways as the
excitation of eccentricity in a nearly circular binary, $J' \propto
m_p/(v_p b^2) $ from Equations \ref{eqDJImpulsive} and
\ref{eqDJNonImpulsive}, it follows that $J' {\cal R}(J') \propto
J'^{-1}$, and $p(J') \propto J'^{-3}$.

The full calculation of $p(J')$ requires choosing the correct
expression for $\Delta \vec J$ given the timescale of the encounters.  In
the extremely non-impulsive regime (Equation \ref{eqDJNonImpulsive}),
$\Delta \vec J$ is averaged over $r_b(t)$ before being used in
Equation \ref{eqRofe}.  For the impulsive case, $\Delta \vec J(t_0)$
retains its dependence on the position of the comet, but the
subsequent integral over $t_0$ in Equation \ref{eqRofe} 
averages the contribution of perturbers from all possible $r_b$.
Ultimately we arrive
at the same $p(J')$ for both non-impulsive and very impulsive
perturbations:

\begin{equation}
\label{eqPStd}
p(J') = 0.74 G \rho a^2 \frac{1}{J'^3},
\end{equation}

\noindent
where $\rho= n m_p$, the volumetric mass density of the perturbers in space.
As noted in CS08, this form of $p(J')$ reveals that the angular
momentum of the comet follows a L\'{e}vy flight \citep{SZF95}. 
The distribution
function is then:

\begin{equation}
\label{eqFofJ}
f(J,t) = \frac{1}{2 \pi J_c^2(t)} (1+(J/J_c(t))^2)^{-3/2}.
\end{equation}

\noindent
This function is self-similar, meaning that it always has the same
shape centered around a characteristic angular momentum scale,
$J_c(t)$, that changes with time.  We have chosen the normalization
such that $\int f(J,t) d^2\vec J=1$ at all times.  The characteristic
angular momentum is near the median of the distribution, $J_{\rm
  median} = \sqrt{3} J_c(t)$.  Since the probability of finding the
comet with an angular momentum of order $J \gg J_c(t)$ falls off like the
power law $J^{-1}$, the mean, variance, and all higher moments of the
distribution are undefined.  The mean only diverges logarithmically;
if there is a maximum angular momentum $J_{\rm max}$, then $J_{\rm
  mean}= 2.3 J_c(t) \log_{10}(0.74 J_{\rm max}/J_c(t))$.

The time derivative of $J_c(t)$ is related to the perturbation 
frequency:

\begin{equation}
\label{eqODE}
\dot J_c(t) = 4.66 G \rho a^2
\end{equation}  

\noindent
This equation is
derived by substituting the solution for $f(J,t)$ (Equation
\ref{eqFofJ}) into the Boltzmann equation (Equation
\ref{eqBoltzmann}).  Equation \ref{eqODE} determines $J_c(t)$ even if
the parameters of the perturbing swarm ($\rho$) or the comet ($a$) 
are changing with time.
During the formation of the 
Oort cloud, the semimajor axes of the comets evolve 
as the ice giants deliver
orbital energy to them over many interactions.
Additionally, a time-varying 
density of perturbers may be relevant if the Sun formed in a 
dense cluster \citep{F97}.  The high eccentricity 
but high periapse orbit of Sedna may imply
that the Sun was born in such an environment \citep{ML04,BDL06,KQ08}.
A realistic statistical description of the formation of the 
Oort cloud must incorporate the evolution of 
$\rho$ and $a$ of the comets.

To provide the following simple numerical example, 
we assume a constant $\rho$ and $a$.
The angular momentum distribution function in this case 
grows linearly with time,
$J_c(t)=4.66G \rho a^2 t$, for $J_c(t) \gg J_c(t=0)$.  Using
values relevant for the Oort cloud, we find

\begin{equation}
\label{eqJcNumbers}
\frac{J_c(t)}{J_{\rm circ}} = 0.363
\left(\frac{\rho}{0.1 M_{\odot} {\rm pc}^{-3}}\right)
\left(\frac{a}{10^4 {\rm AU}}\right)^{3/2}
\left(\frac{t}{1 {\rm Gyr}}\right),
\end{equation}

\noindent
where we have scaled $J_c(t)$ by the angular momentum per unit mass
of a circular orbit, $J_{\rm circ}=\sqrt{G M_{\odot} a}$, to 
make it dimensionless.  Since our derivations neglect 
the non-radial motion of the comet's evolving orbit, our theory is only
quantitatively correct for $J/J_{\rm circ} \ll 1$. 

This mode of growth is qualitatively different from the typical
diffusive random walk.  The passing stars cause a spectrum of
perturbations that occur with frequencies inversely proportional to
their size ($J' {\cal R}(J') \propto J'^{-1}$).  This power law is
such that the smallest kicks cannot accumulate fast enough to affect
the distribution function.  For example, perturbations of about
the same size accumulate as a normal diffusive random walk, $\delta J
\propto \sqrt{t/t_{\rm small}} J'_{\rm small}$.  In that same time,
however, the comet receives, on average, a single perturbation of size
$\delta J \approx J'_{\rm big} \propto (t/t_{\rm small}) J'_{\rm
  small}$.  Thus the overall growth of the angular momentum is due to
the few largest perturbations that occur over a time $t$.

The distribution in angular momentum (Equation \ref{eqFofJ}) 
can be converted to a distribution 
for the comet's periapse distance, $q$, using the relation 
for nearly radial orbits, $J=\sqrt{2 G M_{\odot} q}$:

\begin{equation}
\label{eqFofRp}
f(q,t)=\frac{1}{2 q_c(t)} ( 1+{q}/{q_c(t)})^{-3/2},
\end{equation}

\noindent
where $q_c(t)$ is the characteristic periapse associated with 
$J_c(t)$.  We have chosen a normalization such that $\int f(q,t) d q =
1$.  Since $J_c(t) \propto t$, the typical periapse distance grows as
$t^2$; the timescale for a significant change in periapse then depends
on the comet's current $q$.

These derivations of the distribution of a comet's angular momentum
assumed the swarm of perturbers had a single individual mass 
and single velocity.  If
there are other massive perturbers with $m_p > M_{\odot}$, such as
giant molecular clouds, Equations \ref{eqFofJ} and \ref{eqJcNumbers}
describe the distribution when $\rho$ includes
all of the perturbers: $\rho= \sum{n_i m_{p,i}},$ where $n_i$ and
$m_{p,i}$ are the volumetric number density and masses of the $i$th group
of perturbers.  A mass spectrum that extends significantly below the
mass of the Sun also affects the probability distribution of the 
perturbations.  In the generalized case, the slope of the
perturbation spectrum sets the high $J$ power law of the distribution
function.  As long as the exponent of $J' {\cal R}(J')$ is between $0$
and $-2$, the angular momentum follows a L\'{e}vy flight
\citep{SZF95}.  For the precise details of deriving $p(J')$ and
$f(J',t)$ given a general mass distribution, we refer the reader to
CS08.

\section{Connection to Galactic Tides}
\label{secGalacticTides}

In deriving the model presented in Section \ref{secLevy}, we have
assumed that the perturbing stars are distributed isotropically in
$\hat v_p$ and uniformly in impact parameter.  We expect the angular
momentum distribution in that scenario to be axisymmetric.  Field
stars, which are confined to a disk with a height much less than its
radial dimension, do not have these simplifying properties.  This
section uses a toy model to show how an anisotropy in the angular
momentum distribution arises from the spatial inhomogeneity of the
perturbing stars, and how this is related to the 
angular momentum distribution discussed in section \ref{secLevy}.

\citet{HT86} investigated the effects of the large scale potential
arising from the Galactic disk.  We reproduce their derivation of such
a torque given a simple planar model of the mass distribution.  We
approximate the disk as a stack of infinitely thin, infinitely large
sheets of mass.  Gauss' law shows that the sheets above and below both
the sun and the comet produce no net acceleration on the system.  The
sheets that pass in between the sun and comet however, produce a mean
torque given by:

\begin{equation}
\label{eqGalTorque}
{\dot \vec J} = - 2 \pi G \rho (\vec r_b \cdot \hat z)(\vec r_b \times \hat z),
\end{equation}

\noindent 
where $\rho$ is the local volumetric mass density in perturbers, and
$\hat z$ is the unit vector normal to the disk plane.  To an order of
magnitude, this torque is the same as our Equation \ref{eqODE},
although it is of a completely different nature.  Equation
\ref{eqGalTorque} describes a smooth torque in a fixed direction,
while Equation \ref{eqODE} is the typical value of a stochastic variable
drawn from an axisymmetric distribution with zero mean.

\citet{HT86} also performed numerical
experiments to verify that on very long timescales, stellar scattering
indeed produces a mean growth on top of the stochastic evolution.  The
importance of the Galactic tides has been appreciated in subsequent studies
of Oort cloud dynamics \citep{DQT87,H90,DWLD04,RFFV08}, although the 
relationship between the stellar encounters and the tidal torques is rarely
addressed.
Tidal torques are usually treated as separate
from the effects of stellar encounters, even though the torque is provided by the
same stars that cause the stochastic evolution.  By adapting our
formalism to reflect a planar distribution of perturbers, we reproduce
the effects of the Galactic tides, and in doing so 
find the distribution function that accounts for both modes of 
angular momentum growth.

We follow the example of the numerical experiments of \citet{HT86} and
approximate the Galaxy locally as a uniform disk of material, with a
height much smaller than the scale of the other two dimensions.  To
create the planar symmetry in the model of stellar encounters, the
velocities of the perturbers are restricted to a single direction.
While this is not a realistic representation of the directional
distribution of field star velocities, it is a simple model to explore
and provides a clear example with which to examine the effects
of a velocity asymmetry.  With $\hat v_p$ fixed, the impact parameter
$\vec b$ is confined to a plane, the aspect ratio of which has a much
smaller height than width.  Both of these properties, a single
direction for $\hat v_p$ and a non-unity aspect ratio, introduce
asymmetries in the distribution function of the comet's angular
momentum.

%%!   Note factor of two, and also the third component of the angular momentum

For isotropic perturbers, perturbations of any size $J'$ occur
with the same likelihood in all directions in the plane perpendicular
to $r_b$.
This ensures that the mean of
$\vec J(t)$ is zero, even though the typical magnitude of the angular momentum 
increases linearly with time.  The cross-section for an 
interaction in the tidal limit ($b \gg r_b$) scales as $b^2$, which fixes the 
power law of the single perturbation PDF.  In the planar model, the 
cross-sectional area that contributes perturbations with 
small $J'$ is less than $b^2$ for impact 
parameters larger than the disk height.  
The contributions of these regions to each component of 
$\vec J'$ depends on the angle between the comet and the disk 
plane so the axisymmetry is broken.
However, these differences manifest only in the lowest $J'$, and their 
effects on the distribution of accumulated angular momentum are 
always washed out by the larger perturbations from impact parameters less
than the disk height.

Another asymmetry results from the impact parameters of $b \sim r_b$.  For
$b > r_b$, there is as much cross-sectional area contributing 
positively to each component as there is negatively.  
Impact parameters that pass between the
sun and the comet, however, impart angular momentum in one direction
of one component only, depending on the angle between $\hat r_b$ and
$\hat v_p$.  
Not coincidentally, the
mean torque found in the smooth distribution limit, Equation
\ref{eqGalTorque}, is attributed to the disk of stars passing between
the Sun and the comet.

We quantify the effect of this asymmetry by calculating the marginal probability
density of each component of the angular momentum vector due to
single interactions.  Since we have lost the symmetry that
admitted the simple analytic solutions, we employ a Monte-Carlo
procedure.  The position of the comet, which we hold fixed in this
example, is $\vec r_b = \hat y + \hat z$, so the Sun-comet distance is
$r_b = \sqrt{2}$. The perturber velocities are set to the $\hat z$ direction:
$\hat v_p = -\hat z$.  The possible impact parameters of the perturbers are then
restricted to the $x-y$ plane.  We randomly choose impact parameters
such that they are uniformly distributed over the plane
and calculate the $\Delta \vec J$ delivered to the comet.  We assume
the other parameters of the system are held constant ($v_p$ and
$m_p$), and to reduce the notation, we use units where $2 G m_p/v_p
\equiv 1$.  The angular momentum is confined to the plane
perpendicular to $\vec r_b$, which in these coordinates is defined by
the basis vectors $\hat x$ and $(\hat y - \hat z)/\sqrt{2}$.  For
simplicity we discuss the $x$ and $y$ components of the perturbation, $\Delta \vec
J \cdot \hat x = J'_x$ and $\Delta \vec J \cdot \hat y = J'_y$.  In the
$z$-direction, $\Delta \vec J \cdot \hat z$ is exactly the same as
$J'_y$.  The positive and negative values for $J'_x$ and $J'_y$ are
binned separately; the resulting four histograms then describe the
marginal PDF for each component.

\begin{figure}[t!]
\center
\includegraphics[angle=-90,width=0.75\columnwidth]{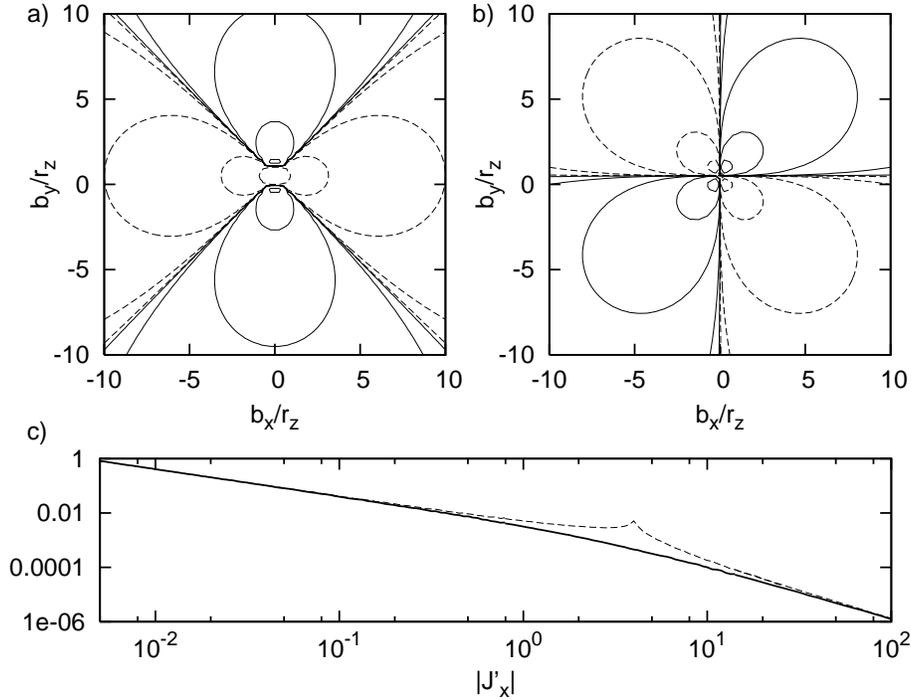}
\caption{Contours of constant $J'$ on the space of impact parameters
$\vec b/(\vec r_b \cdot \hat z)$ for positive and negative values of
each component of the vector perturbation.  The levels are 
spaced in multiples of ten from $J'/(2 G m_p/v_p) = \pm 10^{-4}$ to $\pm 1$. 
Panel b, which shows the contours for the $\hat y$
direction, is symmetric with respect to positive and negative
perturbations.  The center of panel a shows an isolated region of
negative $\hat x$ perturbations that causes an asymmetry in the
distribution function.  By randomly sampling this space of impact
parameters we generate the PDF of the perturbations.  The marginal
PDF for positive and negative $J'_x$ are plotted in panel c; the
spike contains perturbations from the central region of panel a and
is the source of the Galactic tidal torques on the comet.}
\label{figContours}
\end{figure}

Figure \ref{figContours} illustrates the calculation of the single
interaction PDF.  Panels a and b show logarithmically spaced contours
of constant $J'_x$ and $J'_y$ respectively in the plane of possible
impact parameters, with the other parameters of the interaction fixed
($\vec r_b, \vec v_p, m_b$).  The impact parameter plotted is scaled
by $\vec r_b \cdot \hat z = r_z = 1$.  The solid contours correspond to
positive perturbations and the dashed lines to negative ones. In panel
b), the contours for $\pm J'_y$ exhibit an axisymmetric pattern; for
each unit of area that contributes perturbations of a given magnitude
greater than zero, there is an equivalent area where perturbations
have the opposite sign.  Thus the single interaction marginal PDF of
perturbations in the $\hat y$ directions are identical and unchanged
from the isotropic case: ${J'_y}^{-1}$ for the distant perturbations,
$J'_y(b \gg r_b)$, and ${J'_y}^{-2}$ for the close encounters, $J'_y(b
\ll r_b)$.  There is no coherent accumulation of angular momentum in
the $\hat y$ direction.

The contours of panel a), while symmetric at larger $\vec b$, are not
symmetric in the center, where the perturbations only add angular
momentum in the negative $\hat x$ direction.  There is no equivalent
area that delivers angular momentum with the opposite sign.  We plot
the marginal PDF of $J'_x$, $|J'_x| {\cal R}(J'_x)$, in panel c) of
Figure \ref{figContours}, where the solid line is for perturbations
where $J'_x>0$ and the dashed line is for $J'_x<0$.  The values along
the ordinate represent the probability of perturbations with
strength of order $J'_x$ relative to the lowest value plotted.  In the
tidal and close encounter regimes, the two functions are identical.
For $J'_x$ of order unity, the contribution of the central region in
panel a) is obvious.  It is these interactions that give rise to the
torque associated with the Galactic tides.

The marginal PDF of $\vec J'_x$ highlights the source of the Galactic
tidal torque.  However, it remains to describe how this manifests in
the time-dependent distribution function of the comet's angular
momentum.  In Section \ref{secLevy}, we used the Boltzmann equation
(Equation \ref{eqBoltzmann}) to relate the axisymmetric single
perturbation PDF ($p(J')$) to the distribution of angular momentum
($f(\vec J(t))$).  That derivation, however, depends on the
simplifications afforded by the single power law form of $p(J')$.  For
the non-axisymmetric single perturbation PDF depicted in Figure
\ref{figContours}, an analytic solution to the corresponding Boltzmann
equation would be much more difficult to calculate.

%Put another way, the distribution function 
%describes the distribution followed by the
%sum of many single perturbations.  In the isotropic case for low $J'$,
%where the single interaction PDF is a power law, the corresponding
%distribution function for $\vec J(t) = \sum \vec J'_i$ is the
%self-similar two-dimensional Cauchy distribution, equation
%\ref{eqFofJ}.  An anisotropic velocity distribution complicates the 
%solution of the Boltzmann equation.

%--- put this sentiment in the conclusions maybe? --%
%At a literal level, the distribution function for the angular momentum
%is simply the distribution of the sum of many single
%perturbations: $\vec J(t) = \sum \vec J'_i$, where the time 
%$t$ and the frequency of single perturbations dictates the 
%number of terms in this sum.

Instead, we use a bootstrap technique to estimate the distribution
function from a sample of single perturbations.  
The velocity of the perturbers, $v_p$, their number density, 
$n$, and the area sampled when generating the single interaction 
PDF, $\pi b_{\rm max}^2$, set the average time associated with 
each perturbation, $1/\tau = n \pi b_{\rm max}^2 v_p$.  
The angular momentum at a time $t$ is then the sum of 
$t/\tau$ single perturbations.  By randomly choosing $t/\tau$ 
perturbations from the PDF and adding them vectorially, 
we generate a sample of angular momentum vectors that reflect 
the distribution function at that time $t$.

To accurately probe the evolution over many orders of magnitude,
several single interaction PDFs with different $b_{\rm max}$
were used.  Ignoring large impact parameters increases $\tau$, or
equivalently, samples the close encounters more often over a fixed
number of perturbations.  We verified that the distribution functions
calculated with large $\tau$ (small $b_{\rm max}$) are not significantly affected
by ignoring the frequent perturbations of smaller $J'$.

\begin{figure}[t!]
\center
\includegraphics[angle=-90,width=0.75\columnwidth]{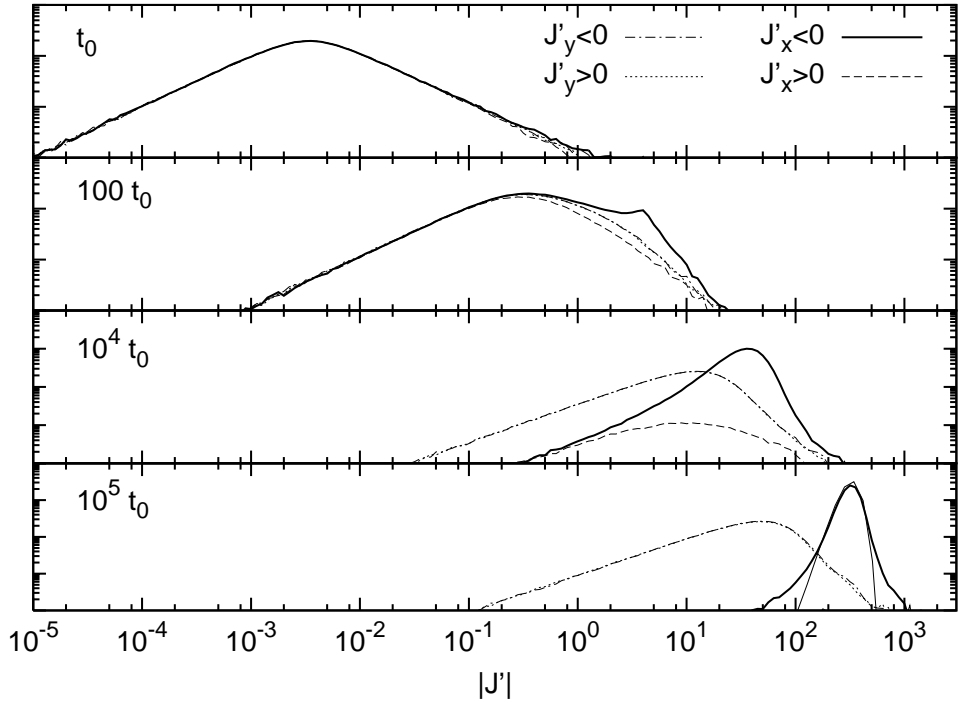}
\caption{Marginal distribution functions of two components of
the angular momentum as a function of time.  The dotted and
dot-dashed lines plot the marginal distribution, $d N/ d (\log
|J_y|)$, of the $\hat y$ component of the angular momentum $\vec
J(t)$.  The thick line is the distribution of the $\hat x$ component
when it is negative, and the dashed line is the positive side.  In
the top two panels, the comet's angular momentum is best described
by the \levy flight behavior caused by stochastic stellar
perturbations.  In the bottom two, the coherent torque attributed to
the Galactic tides dominates the evolution, causing a visibly
asymmetric distribution.  The thin line in the bottom panel is a
Gaussian distribution with the mean given by the Galactic
tidal torques and the variance given by the variance of the single
interaction PDF multiplied by the number of encounters.}
\label{fig4Hists}
\end{figure}

The marginal distribution functions at four different times are shown in Figure
\ref{fig4Hists}.  Each histogram contains $10^6$ bootstrapped $\vec
J(t)$, generated from the sum of between 4 and 1000 single perturbations.  
The distribution of $J_y(t)$ is plotted in the dotted lines for
$J_y(t)>0$ and dash-dotted for $J_y(t)<0$.  For $J_x(t)$, 
the solid line represents the negative perturbations and 
the dashed line the positive ones.

The top panel shows the angular momentum distribution at early times,
or equivalently, at low typical angular momenta.  For reference, we
denote this time $t_0$.  Since the single interaction PDF for perturbations of
this magnitude is axisymmetric, all four functions are identical.   
The excess of perturbations to negative $J'_x$ is not
visible as the likelihood for those encounters is too low to be
sampled in the $10^6$ vectors generated for the plot.

The second panel depicts the four distribution functions 100 times
later than the time of the top panel.  Again both functions show a
similar shape, and the typical value for all four has grown linearly
with time as predicted by Equation \ref{eqODE}.  The trajectories
passing between the sun and the comet have been sampled in a small
fraction of the generated $\vec J(t)$, and the contribution from the
spike of Figure \ref{figContours}c is apparent.  Additionally the
normalization of the positive distribution of $J_x(t)$ has fallen to
reflect the breaking of the symmetry around $J_x=0$.  The
distributions in the first and second panel can be said to be
dominated by the influence of the stellar perturbations, and are not
strongly affected by Galactic tides.  Although the mean of the
distribution is always set by the tides (see Equation
\ref{eqGalTorque}), here this value of angular momentum is only
realized after rare but strong interactions.  The most likely angular
momentum vectors, at early times, are distributed axisymmetrically
around the origin.

In the third panel the non-axisymmetric growth is manifest.  Due to
the higher slope of the single encounter PDF, the distribution of the
$y$ component of the angular momentum has begun to grow only as
$t^{1/2}$; the accumulations of kicks from all of the impact 
parameters smaller than $r_b$ contribute
to the shape of this distribution.  Unfortunately a PDF of this slope
does not admit a self-similar distribution function; asymptotically,
the distribution approaches a Gaussian logarithmically over time \citep{SZF95}.

The perturbers passing between the sun and the comet deliver angular
momentum in the $-\hat x$ direction coherently and thus the typical
$-J_x(t)$ continues to increase linearly in time.  The normalization
of the histogram for positive $J_x(t)$ has decreased substantially,
which is another indicator that the total distribution of $J_x(t)$ is
no longer centered on the origin.  In the fourth panel, only 10 times
later than the third, the marginal distribution function for $J_x(t)$
is entirely dominated by the accumulated effects of non-canceled
encounters.  There are no values of $J_x(t)>0$ in the sample at this
time.  Again, the distribution function does not admit an analytic
form.  For reference, we plot a Gaussian distribution with the mean
described by Equation \ref{eqGalTorque}, and the variance expected
given the single encounter PDF, $\sigma^2 = \sigma_{\rm PDF}^2 t$.
The distribution function only approaches this approximated shape
logarithmically in time.

Figure \ref{fig4Hists} reveals the nature of the coherent torque by
Galactic tides as merely the long term effects of anisotropic stellar
encounters.  It is only a matter of principle what to call the
interactions of the comets with field stars.  To determine the
relevant behavior, one must specify which impact parameters are the
most important for the behavior of the comet.  On shorter timescales,
or for smaller angular momenta, the distant perturbations create the
axisymmetric distribution function associated with stochastic stellar
encounters.  Over timescales long enough that many trajectories have
sampled the region between the Sun and comet, the system is best
characterized as evolving under the Galactic tides.

As a physical example, we again examine the formation of the Oort
cloud, where a proto-comet must gain enough angular momentum to raise
its periapse $q$ by $\Delta q$ to avoid perturbations from the
planets.  The influence of the planets falls off rapidly with
increasing $q$, so a reasonable value for $\Delta q/q$ is on the order
of $10 \%$ \citep{DQT87}.  The distant stellar encounters will be
responsible for building the Oort cloud if a single interaction at an
impact parameter $b \sim a$ can provide enough angular momentum to
increase the periapse.  If these single encounters are too weak, the
coherent growth due to Galactic tides is required.  We find the
following inequality for when the mean tidal growth, rather than
stochastic evolution, dominates:

\begin{equation}
\label{eqOomOort}
\left( \frac{\Delta q}{q} \right) \left( \frac{M_{\odot}}{m_p} \right) 
\left( \frac{v_p}{v_q} \right) \gg 1,
\end{equation}

\noindent
where $v_q = (G m_{\odot} / q)^{1/2}$ is the local rotational velocity at
periapse.  At the semimajor axis of Jupiter, this velocity is about 15
km ${\rm s}^{-1}$, and near Neptune it is about 5 km ${\rm s}^{-1}$.
Typical velocity dispersions of stars in the solar neighborhood are
$15-40~{\rm km~s^{-1}}$ \citep{BT87}.  Then in the inner solar system,
the tidal torque is less important than the stellar encounters for
freeing the comets from planetary perturbations.  In the outer solar
system, the left hand side of Equation \ref{eqOomOort} is close to
unity, meaning the stellar encounters and the tidal torque play a
comparable role.

Our new understanding of the relationship between stellar encounters
and tides presents a clearer picture of the most appropriate way to
model the excitation of angular momentum in an Oort cloud comet.  If
the prescription for stellar encounters includes the planar symmetry
of the stars, then no extra torque is required to represent the
Galactic tides.  If the stellar encounter model has an isotropic
velocity distribution, then an extra term representing the torque
should be included, but only at late enough times that encounters
passing between the sun and the comet are common.

\section{Conclusions}
\label{secConclusions}

In this work we have shown that the angular momentum delivered to
nearly radial comets by passing stars follows a L\'{e}vy flight.  From
the properties of a single scattering between the comet and the star,
we derive the distribution function of the angular momentum of the
comet as a function of time.  Our calculations agree with the
estimates made in earlier work on Oort cloud formation, that stellar
perturbations can raise the periapses of comets significantly in only
several hundred Myrs.  A careful examination of the scattering process
for an anisotropic velocity distribution reveals the presence of the
coherent angular momentum growth that is usually attributed to the
large scale potential of the Galaxy.  The effects of stellar
encounters and the Galactic tidal torques then cannot be treated as
two distinct processes.  On shorter timescales the distribution
function of the comet is unaffected by the tidal torque; on long
timescales the distribution is entirely dominated by it.  Since the
presence of the tidal torque depends on the perturber velocity
distribution, simulations of cometary evolution that include stellar
encounters must be careful not to double-count the Galactic tides by
either including an explicit torque or enforcing a planar symmetry, but 
not both.

These results provide a formal understanding of the effects of stellar
encounters on nearly radial comets, but it is only the first step
towards a complete statistical picture of the formation of the Oort
cloud.  The shape of the distribution function of the angular momentum
at early times will not be entirely isotropic due to the triaxial
velocity distribution of field stars; however, this anisotropy will be
overwhelmed at the current epoch by effects of the Galactic tidal
torque.  The effects of the stellar perturbations must be convolved
with the diffusion of the comets' semimajor axes caused by planetary
perturbations.  This type of diffusion is not without complications,
as orbital resonances between the comet and the planet must be
accounted for to produce accurate diffusion coefficients
\citep{MT99,PS04}.  Additionally, the diffusion of the semimajor axis
for a comet whose orbit crosses that of a planet has been shown to
exhibit properties of a \levy flight \citep{ZSZ02}.

We thank the Institute for Advanced Study for their hospitality
while some of this work was completed.  R.S. is a Packard Fellow.  
This work was partially supported by the ERC.

\bibliographystyle{apj}
\bibliography{ms}

\end{document}